\documentclass[
aps,prl,twocolumn,groupedaddress
]{revtex4-1}

\usepackage{graphicx}
\usepackage{dcolumn}
\usepackage{bm}
\usepackage{natbib}
\usepackage{placeins}
\usepackage{braket,mleftright}
\usepackage{amsmath}
\usepackage{xcolor}

\begin{document}

\preprint{APS/123-QED}

\title{Parity Conservation in Electron-Phonon Scattering in Zigzag Graphene Nanoribbon}

\author{Yanbiao Chu, Pierre Gautreau, Cemal Basaran}
\email{cjb@buffalo.edu}
\affiliation{Electronic Packaging Laboratory, University at Buffalo, SUNY, Buffalo, NY, 14260, USA}

\date{\today}

\begin{abstract}
In contrast with carbon nanotubes, the absence of translational symmetry (or periodical boundary condition) in the restricted direction of zigzag graphene nanoribbon removes the selection rule of subband number conservation. However, zigzag graphene nanoribbons with even dimers do have the inversion symmetry. We, therefore, propose a selection rule of parity  conservation for electron-phonon interactions. The electron-phonon scattering matrix in zigzag graphene nanoribbons is developed using the tight-binging model within the deformation potential approximation. 

\begin{description}

\item[PACS numbers]
72.80.Vp, 73.22.Pr

\end{description}

\end{abstract}

\pacs{Valid PACS appear here}
\maketitle



As one of the most promising materials for future electronics, it is of great importance to understand electrical properties of graphene nanoribbons (GNRs) \cite{son2006half, Chen2007228, neto2009electronic, Lin10062011}. For intrinsic properties like electron mobility, conductance, and resistance, the only contribution comes from phonons \cite{dutta2009intrinsic}. Although graphene nanoribbon can be precisely fabricated by bottom-up approaches \cite{cai2010atomically}, it is still difficult to isolate phonon effects from other external scattering sources such as substrate interactions \cite{PhysRevB.78.205403}. These intrinsic properties must be sought theoretically.

In contrast with carbon nanotubes (CNTs), where the electron/phonon subband numbers can be interpreted as momentum in the circumferrencial direction (which is conserved during scattering), such intepretation does not exist in GNRs.  The electron-phonon scattering in armchair graphene nanoribbons (AGNRs) has been studied extensively \cite{PhysRevB.78.205403, Betti2007APL, PhononLimited2012APL, ModeSpace2013APL}. This is, however, not the case for zigzag graphene nanoribbons (ZGNRs). According to Betti et al.  \cite{Betti2007APL}, electrons in AGNRs can be scattered by phonon of any subband and jump to any final electron subband, without following the strict subband selection rule of CNTs. This phenomenon is called transverse momentum conservation uncertainty. 
 
In this study, however, we propose to significantly mitigate this uncertainty  by introducing parity, a concept related to the inversion or mirror symmetry of ZGNR with even dimers. The mirror symmetry itself is not a new concept for ZGNRs. $Ab \ initio$ computations generate distinctly different transport behavior for ZGNRs with and without mirror symmetry about their central line \cite{PhysRevLett.100.206802}. When scattered by a steplike barrier, electrons in ZGNRs preserve the parity of the wave function expressed by the recursive Green's function method \cite{PhysRevLett.102.066803}. In this study, parity conservation is shown to stand for electron-phonon interactions as well. 

Intuitively, this generalization is quite natrual, as the conservation of a physical quantity is equivalent to the commutative of its corresponding operator with the Hamiltonian. ZGNRs with even dimers do have mirror symmetry about their central line and the Hamiltonian doesn't change under inversion operation. The parity conservation will become apparent in deriving the electron-phonon scattering matrix elements. During this derivation, the only approximations that were made are ones that do not harm the parity of the system, and allow us to demonstrate parity conservation.

\begin{figure}[h]
\centering
\includegraphics[scale=0.45]{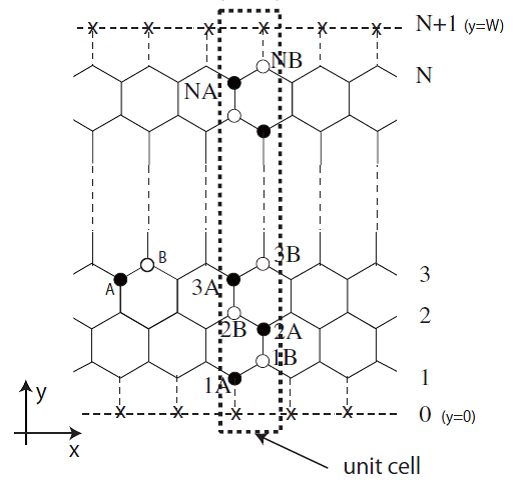}
\caption{\label{fig:ZGNR} Lattice of ZGNR. The unit cell of ZGNR is the vertical atomic chain within the dashed rectangle. Following the routine of graphene, atoms in the unit cell are classified into either A type or B type. Each pair of A-B carbon atoms are called a dimer.  Adapted from \cite{1468-6996-11-5-054504}}
\end{figure}

First, let's study the parity of electron state in ZGNR.  The lattice of ZGNR is shown in Fig.~\ref{fig:ZGNR}, with the $sp^2$ hybridized carbon atoms on the edges passivated by hydrogen atoms. Hence, the probability for the formation of $\pi$-electrons bond on the edge sites denoted by $\times$ is zero. In other words, the edges of the nanoribbon are treated as hard walls for electrons, where the wave function should be zero.

In the single-electron nearest-neighbor tight-binding model of graphene, the $\pi$-electron at $R_A$ site of an A-sublattice can be created by operator $\hat{\alpha}_{R_A}^\dagger$ or annihilated by operator $\hat{\alpha}_{R_A}$. Identically for $R_B$ site, we define $\hat{\beta}_{R_B}^\dagger$ and $\hat{\beta}_{R_B}$ as the creation and annihilation operators.  With the hopping parameter $t$, the Hamiltonian of ZGNR can be written as 
\begin{equation} \begin{split}
H=&-t\sum\limits_{l}\sum\limits_{d=1}^N\left[\hat{\alpha}_l^\dagger(d)\hat{\beta}_{l-1}(d) + \hat{\beta}_l^\dagger(d)\hat{\alpha}_l(d)\right]\\
& + H.c.-t\sum\limits_{l}\sum\limits_{d=1}^{N-1}\hat{\alpha}_l^\dagger(d+1)\hat{\beta}_{l}(d)+H.c.
\end{split} \end{equation}
where $l$ is the unit cell index, $d$ is the dimer index, and $H.c.$ is the hermitian conjugate of the preceding term.

This problem can be solved analytically for both energy bands and electron states \cite{1468-6996-11-5-054504}. For ZGNRs with N dimers, the wave function of extended Bloch state $\ket{\Psi(k)}$ can be written as
\begin{equation}
\begin{split}
\ket{\Psi(k)}&=\sum_d\left( \psi_{d,A}\sum_l\left( e^{ikx_{l,d,A}}\hat{\alpha}_l^\dagger(d)\right)\right)\ket 0\\
&+\sum_d \left( \psi_{d,B}\sum_l\left(e^{ikx_{l,d,B}}\hat{\beta}_l^\dagger(d)\right)\right)\ket 0
\end{split}
\end{equation}
where $\psi_{d,A}=\sin[\xi(N+1-d)]$, $\psi_{d,B}=\pm\sin(\xi d)$, $\xi$ is the quantization parameter in the $y$ direction; $k$ is the wavenumber in the $x$ direction, and $x_{l,d,A}$ ($x_{l,d,B}$) is the $x$ coordinate of A (B) type carbon atom in dimer $d$ of unit cell $l$. 

The energy $E$ of electron states  with $\psi_{d,B}=\sin(\xi d)$ can be expressed as
\begin{equation}
\begin{split}
E=&\left\{g_k\cos(\xi(N+1))  +\cos(\xi N)\right\}\\
&+i\left\{g_k\sin(\xi(N+1)) + \sin(\xi N)\right\},
\end{split}
\end{equation}
where $g_k=2\cos(k/2)$. Since energy $E$ is a real number, the imaginary part should always be zero. Thus,
\begin{equation}
F(\xi,N)\equiv g_k\sin(\xi(N+1)) + \sin(\xi N)=0,
\end{equation}
where the quantized $\xi$ for each subband is solved as root of $F(\xi,N)$ and depends on the longitudinal wave number $k$. Taking the derivative of $F(\xi,N)$ with respect to $\xi$, we'll get
\begin{equation}
\begin{split}
F'(\xi,N)&= (N+1)g_k\cos(\xi(N+1)) + N\cos(\xi N)\\
&\approx (N+1)\left\{g_k\cos(\xi(N+1))  +\cos(\xi N)\right\}\\
&=(N+1)E
\end{split}.
\end{equation}

As $g_k$ is positive for all wave number $k$ within the first Brillouin zone, $F(\xi,N)$ should be also postive beginning with $\xi=0$ . The first root of $F(\xi,N)$ happens when it crosses the horizontal axis from positive side to negative side. Therefore, $F'(\xi,N)$ at the first root is negative. Following the same logic, $F'(\xi,N)$  at next root is positve, then negative, and so on. Equivalently, the energy for electron states with $\psi_{d,B}=\sin(\xi d)$ is alternatively negative, positive, negative, $\ldots$ In other words, it corresponds to alternating pattern of valence band, conduction band, valence band, and so on. Similarly, for electron states with $\psi_{d,B}=-\sin(\xi d)$, the corresponding energy is postive, negative, positve, $\ldots$ Therefore, the conduction bands with increasing value $\xi$ have alternatively electron state with $\psi_{d,B}=-\sin(\xi d)$ and $\psi_{d,B}=\sin(\xi d)$.

As shown in Fig.~\ref{fig:ZGNR}, the nomenclature $(d,A)$ and $(d,B)$ are generated by labelling each carbon atom in the unit cell from the bottom up. If the atoms are labelled inversely from top to bottom, then A type atom $(d,A)$ will have an index of $(N+1-d,B)$. In other words, $\psi_{d,A}$ in one nomenclature would be $\psi_{N+1-d,B}$ in another nomenclature, and vice versa. Since $\psi_{N+1-d,B}=\pm\sin(\xi(N+1-d))=\pm\psi_{d,A}$, thus the electron state with $+$ sign of $\psi_{d,B}$ has even parity while the electron state with $-$ sign of $\psi_{d,B}$ has odd parity. 

\begin{figure}[h]
\hbox{\hspace{0cm}\includegraphics[scale=0.3]{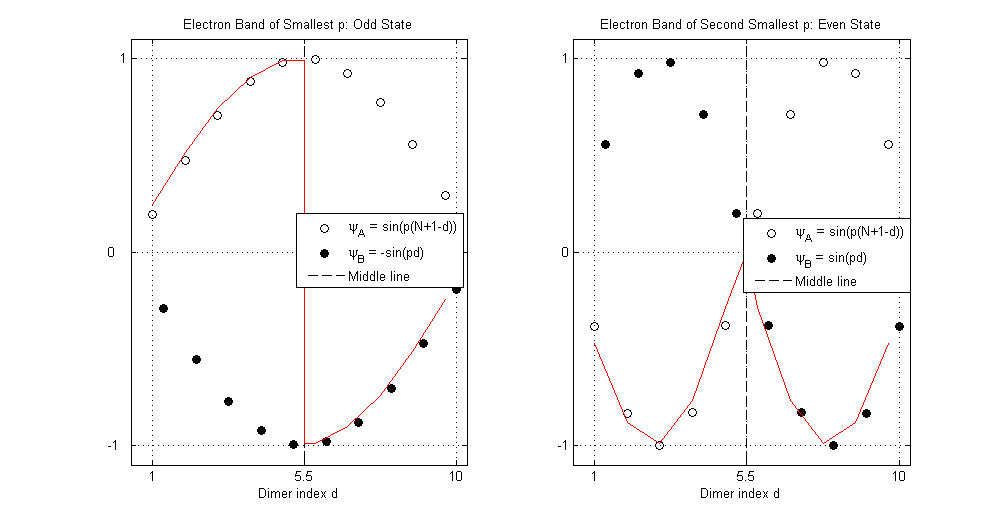}}
\caption{\label{fig:Electron} For ZGNRs with 10 dimers, $\psi_{m,A}$ (open circle) and $\psi_{m,B}$ (solid circle) for the first and second electron bands. Solid line is the average between the absolute magnitude of $\psi_{m,A}$ and $\psi_{m,B}$.}
\end{figure}

Based on the preceding discussion, the parity of each conduction band with $\xi$ from smallest to largest is alternatively odd, even, odd, and so on. Individually, neither $\psi_{d,A}$ nor $\psi_{d,B}$ have even or odd attributes about the central line as shown in Fig.~\ref{fig:Electron}. The parity is demonstrated by the relationship between $\psi_{d,A}$ and $\psi_{N+1-d,B}$. 
The observation that can be made is, for conduction bands, the signs of $\psi_{d,A}$ and $\psi_{d,B}$ for each dimer $d$ are always opposite to each other. It's analogus to the $H_2^+$ ion situation, where the electron state with two out-phase components (anti-bonding state) always has higher energy than the electron state with two in-phase components (bonding state). Similar phenomeona can also be observed for electron states of graphene, as it has two atoms within one unit cell. 

If we approximately treat each dimer of ZGNR as one unit cell of graphene, then the out-phase between $\psi_{d,A}$ and $\psi_{d,B}$ will be taken care of automatically. We only need to describe the average magnitude $\psi_{d,ave}$ for each dimer. Considering the out-phase and the parity,  it should be calculated as $(\psi_{d,A}-\psi_{d,B})/2$ for dimers $d\leq N/2$ and $(\psi_{d,B}-\psi_{d,A})/2$ for dimers $d>N/2$. 
\begin{equation}
\begin{split}
\psi_{d,ave}&=\pm\frac{\psi_{d,A}-\psi_{d,B}}{2}\\
&=\pm\cos\frac{\xi(N+1)}{2}\sin\frac{\xi(N+1-2d)}{2}
\end{split}
\end{equation} 
Such averages are plotted as the red lines in Fig.~\ref{fig:Electron} and present a better demonstration of the parity of electron state. It can be normalized by constant $N_e=\sum_d \cos^2\frac{\xi(N+1)}{2}\sin^2\frac{\xi(N+1-2d)}{2}$.

\begin{figure}[h]
\centering
\includegraphics[scale=0.45]{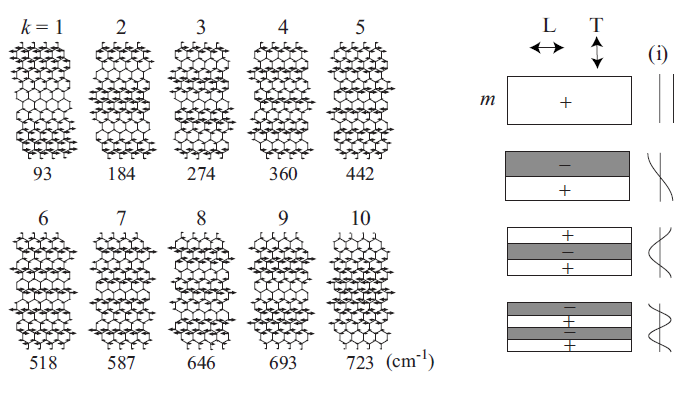}
\caption{\label{fig:Phonon} Longitudinal acoustic modes of ZGNR with 10 dimers, at longitudinal lattice wave number $q=0$. $k$ is the number of nodes in width direction.   Adapted from \cite{PhysRevB.77.054302}}
\end{figure}

For the parity of phonons in ZGNRs, however, we have to rely on some phenomenological arguments. As although lattice vibration is enssentially a problem of classical mechanics, analytical solutions for phonon wavefunctions are still rarely available. In contrast, various computations have been done for ZGNRs by ab initio method\cite{PhysRevB.77.054302,PhysRevB.80.155418}, and other methods. All these analyses generate similar results that the normal modes of ZGNR have a nodal structure in the width direction, as shown in Fig.~\ref{fig:Phonon}. 

Similar to the vibration of membranes with two free edges, we could expect that the modal shapes have the forms of sine functions (odd) or cosine functions (even). Approximately, the wave length $\lambda$ and wave number $\eta$ of mode $p$ can be expressed as
\begin{equation}
\lambda_p=\frac{2W}{k}, \qquad \eta_p=\frac{2\pi}{\lambda}=\frac{k}{W}\pi,
\end{equation}
where $W$ is the width of ZGNR and $k$ is the number of nodes in the width direction. Therefore, the magnitude of polarization $\mathbf{e}_{q,b,p}$ of mode $p$ with longitudinal wave number $q$ at basis $b$ can be approximated by
\begin{equation}
 ||\mathbf{e}_{q,b,p}||= 
\begin{cases}
    \cos(\frac{\eta(N+1-2d)}{2}), & \text{even mode},\\
    \sin(\frac{\eta(N+1-2d)}{2}), & \text{odd mode},
  \end{cases}
\end{equation}
where the polyatomic basis $b$ could be $\{d, A\}$ or $\{d, B\}$. Again, here we approximate one dimer of ZGNR by the unit cell of graphene and only describe the average magnitude of polarization. Respectively, they should be normalized by $N_{ph}^e=\sum_d\cos^2(\frac{\eta(N+1-2d)}{2})$ for even modes and $N_{ph}^o=\sum_d\sin^2(\frac{\eta(N+1-2d)}{2})$ for odd modes.

As quanta of the lattice vibration, phonons are related to the ion displacements $\mathbf{q}_{l,b}$ by their creation (annihilation) operator $a_{q,p}^\dagger$ ($a_{q,p}$) as \cite{ziman}
\begin{equation}\label{eq5}
\mathbf{q}_{l,b}=-i\sum\limits_{q,p}\sqrt{\frac{\hbar}{2Nm_b\nu_{q,p}}}e^{iql}\mathbf{e}_{q,b,p}(a_{q,p}^\dagger - a_{-q,p}),
\end{equation}
where $m_b$ is the mass of atom $b$ and $N$ is the number of atoms in the system. Thus, 
the perturbation of the potential $\mathcal{U}$ between electron and ion lattice is
\begin{equation}
\delta\mathcal{U} = \sum\limits_{l,b}{\mathbf{q}_{l,b}}\frac{\partial \mathcal{U}(\mathbf{r}_i)}{\partial {\mathbf{q}_{l,b}}},
\end{equation}
where $\mathbf{r}_i$ is the coordinate of electron. 

According to Fermi's Golden Rule, we only need $\mathcal{M}({k,k'})$ to calculate the probability for an electron gets scattered by a specific phonon mode from intial state $k$ to final state $k'$, 
\begin{equation}\label{Mkk}
\begin{split}
&\mathcal{M}({k,k'})  = \left< n_{q,p},\Psi_{k'}^* | \delta\mathcal{U}  | n_{q,p}-1,\Psi_{k}\right>, \\
&=i \sum\limits_{\mathbf{l},b}\sqrt{\frac{\hbar n_{q,p}}{2N m \nu_q}} e^{ilq}{\bra{\Psi_{k'}}\mathbf e_{k,b,p}\frac{\partial \mathcal{U}(\mathbf{r}_i)}{\partial {\mathbf{q}_{\mathbf{l},b}}}\ket{\Psi_k}}.
\end{split}
\end{equation}

The kernel part $\bra{\Psi_{k'}}\mathbf e_{k,b,p}\frac{\partial \mathcal{U}(\mathbf{r}_i)}{\partial {\mathbf{q}_{\mathbf{l},b}}}\ket{\Psi_k}$ can be worked out more explicitly by plugging in the analytical solution of the electron states, as
\begin{equation}
\begin{split}
&{\bra{\Psi_{k'}}\mathbf e_{k,b,p}\frac{\partial \mathcal{U}(\mathbf{r}_i)}{\partial {\mathbf{q}_{\mathbf{l},b}}}\ket{\Psi_k}}\\
=&{\sum_{d',l'}\sum_{d",l"}}\bra{0}(\psi_{d',A}{e^{-ik'x_{l',d'A}}}\hat{\alpha}_{l'}+\psi_{d',B}\\
&\times{e^{-ik'x_{l',d'B}}}\hat{\beta}_{l'})\cdot\left(\mathbf e_{k,b,p}\frac{\partial \mathcal{U}(\mathbf{r}_i)}{\partial {\mathbf{q}_{\mathbf{l},b}}}\right)\cdot(\psi_{d",A}\\
&\times{e^{ikx_{l'',d''A}}}\hat{\alpha}_{l''}^\dagger(d'')+\psi_{d'',B}{e^{ikx_{l'',d''B}}}\hat{\beta}_{l''}^\dagger(d''))\ket{0}
\end{split}.
\end{equation}

The two sums on the right hand side can be dropped according to the following arguments. First,  ${\partial \mathcal{U}(\mathbf{r}_i)}/{\partial {\mathbf{q}_{\mathbf{l},b}}}$ describes the change of potential only due to the displacement of atom at site $\{l,b\}$. It reasonable to expect that this change is quite localized and has negligible magnitude at other sites. Second, the whole perturbation term $\mathbf e_{k,b,p} \cdot {\partial \mathcal{U}(\mathbf{r}_i)}/{\partial {\mathbf{q}_{\mathbf{l},b}}}$ is treated as a constant within the deformation potential approximation and  can be factored out as a constant. Finally, according to the assumptions of  the tight binding model, the creation and annihilation operators of electron $\hat \alpha_{l}(m)^\dagger$, $\hat \beta_{l}(m)^\dagger$, $\hat \alpha_{l}(m)$, $\hat \beta_{l}(m)^\dagger$ are orthogonal to each other between different sites. Therefore, only terms with $l=l'=l''$ and $d=d'=d''$ are non-zero.

 Following the preceding arguments and applying the Fermion's anticommutation relations, the kernel part $\bra{\Psi_{k'}}\mathbf e_{k,b,p}\frac{\partial \mathcal{U}(\mathbf{r}_i)}{\partial {\mathbf{q}_{\mathbf{l},b}}}\ket{\Psi_k}$ can be rewritten as
\begin{equation}
\begin{split}
&{\bra{\Psi_{k'}}\mathbf e_{k,b,p}\frac{\partial \mathcal{U}(\mathbf{r}_i)}{\partial {\mathbf{q}_{\mathbf{l},b}}}\ket{\Psi_k}}\\
&=\mathbf e_{k,dA,p}\frac{\partial \mathcal{U}(\mathbf{r}_i)}{\partial {\mathbf{q}_{\mathbf{l},dA}}}\psi_{d,A}(k')\psi_{d,A}(k){e^{i(k-k')x_{l,dA}}}\\
&+\mathbf e_{k,dB,p}\frac{\partial \mathcal{U}(\mathbf{r}_i)}{\partial {\mathbf{q}_{\mathbf{l},dB}}}\psi_{d,B}(k')\psi_{d,B}(k){e^{i(k-k')x_{l,dB}}}
\end{split}.
\end{equation}
The plane wave term $e^{i(k-k')x_{l,dA}}$ can be factored out as $e^{i(k-k')l}$, where $l=x_{l,dA}$. The remaining terms in $e^{i(k-k')x_{l,dB}}$ become $e^{i(k-k')(x_{l,dA}-x_{l,dB})}=\theta_{kk'}$. 

Since $e^{i(k-k'+q){l}}$ is constant for all basis $b$ in unit cell $l$, it can be taken out. Therefore, the summation in the matrix elements $\mathcal{M}({k,k'})$ over $l$ and $d$ can be broken into two separate parts as,
\begin{equation}\label{Mkk}
\mathcal{M}({k,k'})=i \sqrt{\frac{\hbar n_{q,p}}{2N m \nu_q}} \sum\limits_l e^{i(k-k'-q)l}\mathcal{I}_q({k,k'}),
\end{equation}
where the factor $\mathcal{I}_q({k,k'})$ is a summation over dimers $d$ as
\begin{equation}
\begin{split}
&\mathcal{I}_q({k,k'})=\sum_d\Large(\mathbf e_{k,dA,p}\frac{\partial \mathcal{U}(\mathbf{r}_i)}{\partial {\mathbf{q}_{\mathbf{l},dA}}}\psi_{d,A}(k')\psi_{d,A}(k)\\
&+\mathbf e_{k,dB,p}\frac{\partial \mathcal{U}(\mathbf{r}_i)}{\partial {\mathbf{q}_{\mathbf{l},dB}}}\psi_{d,B}(k')\psi_{d,B}(k) e^{i\theta_{kk'}})\equiv \sum_d \Phi
\end{split}
\label{Ieq}
\end{equation}

With a proper normalization of the electron wave function, the summation $\sum\limits_le^{-i(k-k'-q)l}$ yields a Kronecker delta $\delta_{g,(k-k'-q)}$, where $g$ is a reciprocal lattice constant of the ZGNR. This can be interpreted as the conservation of crystal momentum in the longitudinal direction,
\begin{equation}
q=k-k'-g,
\end{equation}
up to an arbitrary reciprocal lattice value $g$.

The overlap term $\mathcal{I}_q({k,k'})$ is central to the electron-phonon interaction, which is a summation over the multiplication of three components with parity: phonon polarization vectors $\mathbf e_{k,b,p}$ , initial electron wave coefficients $\psi_d$, and final electron wave coefficients $\psi'_d$. As an operator, parity only has two eigenvalues of $+1$ (even) and $-1$ (odd). Obviously, even$\times$even = even, odd$\times$odd = odd, odd$\times$even=odd. For the three-particle process here, it's equivalent to: 
\begin{equation}
Parity(p_1)\times Parity(p_2)\times Parity(p_3)=even,
\end{equation}
where $p_i$ represents either a phonon or electron. Therefore, the summation $\mathcal{I}_q({k,k'})$ is non-zero only when $\Phi$ is even about the dimer index $d$. Acoording to this proposed parity selection rule, $\mathcal{I}_q({k,k'})$ can be calculated using half of the dimers, i.e. $\sum_d \Phi=\sum_{d=1}^{N/2} \Phi$. 

Additionaly, as $\Phi$ describes the phonon-electron interaction within a dimer, a good approximation can be made by using the deformation potential of graphene,
\begin{equation}
\Phi  \propto i \sqrt{\frac{\hbar n_{q,p}}{2N m \nu_q}}D(1+e^{i\theta_{kk'}}).
\end{equation}
The proportional constant is due to the differences of normalization schemes for both electron and phonon between ZGNR and graphene.

For the scattering process with electron jumping from even state to even state or odd state to odd state, the parity of the involved phonon can only be even. Similarly, only odd phonons can scatter electrons from even state to odd state or odd state to even state. The matrix element $\mathcal{M}({k,k'})$ will be
\begin{equation}
\begin{split}
 &\mathcal{M}({k,k'})= \delta_{g,k+q-k'}2\sum_{d=1}^{N/2}  i \sqrt{\frac{\hbar n_{q,p}}{2N m \nu_q}}D(1+e^{i\theta_{kk'}})    \\
&\times \frac{\psi_{d}}{N_e} \frac{\psi'_d}{N'_e}
\begin{cases}
    \cos(\frac{\eta(N+1-2d)}{2})/N_{ph}^e, & \text{even phonon}.\\
    \sin(\frac{\eta(N+1-2d)}{2})/N_{ph}^o, & \text{odd phonon}.
  \end{cases}
\end{split}
\end{equation}

Different than the transverse momentum conservation uncertainty proposed by Ref. \cite{Betti2007APL}, which claimed the absence of a selection rule, we have demonstrated that parity can be used as a selection rule for electron-phonon scattering events in ZGNRs with even dimers. Since parity conservation is related with the mirror symmetry of the lattice about its central line, the results presented can be extended to AGNRs due to AGNRs also having the mirror symmetry. It is worth noting that, the mirror symmetry for electrons can be easily destroyed by applying an electrical field in the transverse direction. Then the selection rule of parity conservation breaks down and many more scattering mechanisms are available to electrons. Therefore, for both AGNRs and ZGNRs, we should expect different electron transport behaviors with and without the presence of an applied tranverse electric field.

We would like to recognize the contribution of Dr. Xuedong Hu of the Physics Department at the University at Buffalo for our insightful discussions. We also gratefully acknowledge the financial support received from the US Navy Office of Naval Research Advanced Electrical Power Systems program, under the direction of Dr. Peter Chu. 

\bibliographystyle{apsrev4-1}
\bibliography{Parity}

\begin{thebibliography}{16}%
\makeatletter
\providecommand \@ifxundefined [1]{%
 \@ifx{#1\undefined}
}%
\providecommand \@ifnum [1]{%
 \ifnum #1\expandafter \@firstoftwo
 \else \expandafter \@secondoftwo
 \fi
}%
\providecommand \@ifx [1]{%
 \ifx #1\expandafter \@firstoftwo
 \else \expandafter \@secondoftwo
 \fi
}%
\providecommand \natexlab [1]{#1}%
\providecommand \enquote  [1]{``#1''}%
\providecommand \bibnamefont  [1]{#1}%
\providecommand \bibfnamefont [1]{#1}%
\providecommand \citenamefont [1]{#1}%
\providecommand \href@noop [0]{\@secondoftwo}%
\providecommand \href [0]{\begingroup \@sanitize@url \@href}%
\providecommand \@href[1]{\@@startlink{#1}\@@href}%
\providecommand \@@href[1]{\endgroup#1\@@endlink}%
\providecommand \@sanitize@url [0]{\catcode `\\12\catcode `\$12\catcode
  `\&12\catcode `\#12\catcode `\^12\catcode `\_12\catcode `\%12\relax}%
\providecommand \@@startlink[1]{}%
\providecommand \@@endlink[0]{}%
\providecommand \url  [0]{\begingroup\@sanitize@url \@url }%
\providecommand \@url [1]{\endgroup\@href {#1}{\urlprefix }}%
\providecommand \urlprefix  [0]{URL }%
\providecommand \Eprint [0]{\href }%
\providecommand \doibase [0]{http://dx.doi.org/}%
\providecommand \selectlanguage [0]{\@gobble}%
\providecommand \bibinfo  [0]{\@secondoftwo}%
\providecommand \bibfield  [0]{\@secondoftwo}%
\providecommand \translation [1]{[#1]}%
\providecommand \BibitemOpen [0]{}%
\providecommand \bibitemStop [0]{}%
\providecommand \bibitemNoStop [0]{.\EOS\space}%
\providecommand \EOS [0]{\spacefactor3000\relax}%
\providecommand \BibitemShut  [1]{\csname bibitem#1\endcsname}%
\let\auto@bib@innerbib\@empty
\bibitem [{\citenamefont {Son}\ \emph {et~al.}(2006)\citenamefont {Son},
  \citenamefont {Cohen},\ and\ \citenamefont {Louie}}]{son2006half}%
  \BibitemOpen
  \bibfield  {author} {\bibinfo {author} {\bibfnamefont {Y.-W.}\ \bibnamefont
  {Son}}, \bibinfo {author} {\bibfnamefont {M.~L.}\ \bibnamefont {Cohen}}, \
  and\ \bibinfo {author} {\bibfnamefont {S.~G.}\ \bibnamefont {Louie}},\
  }\href@noop {} {\bibfield  {journal} {\bibinfo  {journal} {Nature}\ }\textbf
  {\bibinfo {volume} {444}},\ \bibinfo {pages} {347} (\bibinfo {year}
  {2006})}\BibitemShut {NoStop}%
\bibitem [{\citenamefont {Chen}\ \emph {et~al.}(2007)\citenamefont {Chen},
  \citenamefont {Lin}, \citenamefont {Rooks},\ and\ \citenamefont
  {Avouris}}]{Chen2007228}%
  \BibitemOpen
  \bibfield  {author} {\bibinfo {author} {\bibfnamefont {Z.}~\bibnamefont
  {Chen}}, \bibinfo {author} {\bibfnamefont {Y.-M.}\ \bibnamefont {Lin}},
  \bibinfo {author} {\bibfnamefont {M.~J.}\ \bibnamefont {Rooks}}, \ and\
  \bibinfo {author} {\bibfnamefont {P.}~\bibnamefont {Avouris}},\ }\href
  {\doibase http://dx.doi.org/10.1016/j.physe.2007.06.020} {\bibfield
  {journal} {\bibinfo  {journal} {Physica E: Low-dimensional Systems and
  Nanostructures}\ }\textbf {\bibinfo {volume} {40}},\ \bibinfo {pages} {228 }
  (\bibinfo {year} {2007})},\ \bibinfo {note} {international Symposium on
  Nanometer-Scale Quantum Physics}\BibitemShut {NoStop}%
\bibitem [{\citenamefont {Neto}\ \emph {et~al.}(2009)\citenamefont {Neto},
  \citenamefont {Guinea}, \citenamefont {Peres}, \citenamefont {Novoselov},\
  and\ \citenamefont {Geim}}]{neto2009electronic}%
  \BibitemOpen
  \bibfield  {author} {\bibinfo {author} {\bibfnamefont {A.~C.}\ \bibnamefont
  {Neto}}, \bibinfo {author} {\bibfnamefont {F.}~\bibnamefont {Guinea}},
  \bibinfo {author} {\bibfnamefont {N.}~\bibnamefont {Peres}}, \bibinfo
  {author} {\bibfnamefont {K.~S.}\ \bibnamefont {Novoselov}}, \ and\ \bibinfo
  {author} {\bibfnamefont {A.~K.}\ \bibnamefont {Geim}},\ }\href@noop {}
  {\bibfield  {journal} {\bibinfo  {journal} {Reviews of modern physics}\
  }\textbf {\bibinfo {volume} {81}},\ \bibinfo {pages} {109} (\bibinfo {year}
  {2009})}\BibitemShut {NoStop}%
\bibitem [{\citenamefont {Lin}\ \emph {et~al.}(2011)\citenamefont {Lin},
  \citenamefont {Valdes-Garcia}, \citenamefont {Han}, \citenamefont {Farmer},
  \citenamefont {Meric}, \citenamefont {Sun}, \citenamefont {Wu}, \citenamefont
  {Dimitrakopoulos}, \citenamefont {Grill}, \citenamefont {Avouris},\ and\
  \citenamefont {Jenkins}}]{Lin10062011}%
  \BibitemOpen
  \bibfield  {author} {\bibinfo {author} {\bibfnamefont {Y.-M.}\ \bibnamefont
  {Lin}}, \bibinfo {author} {\bibfnamefont {A.}~\bibnamefont {Valdes-Garcia}},
  \bibinfo {author} {\bibfnamefont {S.-J.}\ \bibnamefont {Han}}, \bibinfo
  {author} {\bibfnamefont {D.~B.}\ \bibnamefont {Farmer}}, \bibinfo {author}
  {\bibfnamefont {I.}~\bibnamefont {Meric}}, \bibinfo {author} {\bibfnamefont
  {Y.}~\bibnamefont {Sun}}, \bibinfo {author} {\bibfnamefont {Y.}~\bibnamefont
  {Wu}}, \bibinfo {author} {\bibfnamefont {C.}~\bibnamefont {Dimitrakopoulos}},
  \bibinfo {author} {\bibfnamefont {A.}~\bibnamefont {Grill}}, \bibinfo
  {author} {\bibfnamefont {P.}~\bibnamefont {Avouris}}, \ and\ \bibinfo
  {author} {\bibfnamefont {K.~A.}\ \bibnamefont {Jenkins}},\ }\href {\doibase
  10.1126/science.1204428} {\bibfield  {journal} {\bibinfo  {journal}
  {Science}\ }\textbf {\bibinfo {volume} {332}},\ \bibinfo {pages} {1294}
  (\bibinfo {year} {2011})}\BibitemShut {NoStop}%
\bibitem [{\citenamefont {Dutta}\ \emph {et~al.}(2009)\citenamefont {Dutta},
  \citenamefont {Manna},\ and\ \citenamefont {Pati}}]{dutta2009intrinsic}%
  \BibitemOpen
  \bibfield  {author} {\bibinfo {author} {\bibfnamefont {S.}~\bibnamefont
  {Dutta}}, \bibinfo {author} {\bibfnamefont {A.~K.}\ \bibnamefont {Manna}}, \
  and\ \bibinfo {author} {\bibfnamefont {S.~K.}\ \bibnamefont {Pati}},\
  }\href@noop {} {\bibfield  {journal} {\bibinfo  {journal} {Physical review
  letters}\ }\textbf {\bibinfo {volume} {102}},\ \bibinfo {pages} {096601}
  (\bibinfo {year} {2009})}\BibitemShut {NoStop}%
\bibitem [{\citenamefont {Cai}\ \emph {et~al.}(2010)\citenamefont {Cai},
  \citenamefont {Ruffieux}, \citenamefont {Jaafar}, \citenamefont {Bieri},
  \citenamefont {Braun}, \citenamefont {Blankenburg}, \citenamefont {Muoth},
  \citenamefont {Seitsonen}, \citenamefont {Saleh}, \citenamefont {Feng} \emph
  {et~al.}}]{cai2010atomically}%
  \BibitemOpen
  \bibfield  {author} {\bibinfo {author} {\bibfnamefont {J.}~\bibnamefont
  {Cai}}, \bibinfo {author} {\bibfnamefont {P.}~\bibnamefont {Ruffieux}},
  \bibinfo {author} {\bibfnamefont {R.}~\bibnamefont {Jaafar}}, \bibinfo
  {author} {\bibfnamefont {M.}~\bibnamefont {Bieri}}, \bibinfo {author}
  {\bibfnamefont {T.}~\bibnamefont {Braun}}, \bibinfo {author} {\bibfnamefont
  {S.}~\bibnamefont {Blankenburg}}, \bibinfo {author} {\bibfnamefont
  {M.}~\bibnamefont {Muoth}}, \bibinfo {author} {\bibfnamefont {A.~P.}\
  \bibnamefont {Seitsonen}}, \bibinfo {author} {\bibfnamefont {M.}~\bibnamefont
  {Saleh}}, \bibinfo {author} {\bibfnamefont {X.}~\bibnamefont {Feng}},  \emph
  {et~al.},\ }\href@noop {} {\bibfield  {journal} {\bibinfo  {journal}
  {Nature}\ }\textbf {\bibinfo {volume} {466}},\ \bibinfo {pages} {470}
  (\bibinfo {year} {2010})}\BibitemShut {NoStop}%
\bibitem [{\citenamefont {Fang}\ \emph {et~al.}(2008)\citenamefont {Fang},
  \citenamefont {Konar}, \citenamefont {Xing},\ and\ \citenamefont
  {Jena}}]{PhysRevB.78.205403}%
  \BibitemOpen
  \bibfield  {author} {\bibinfo {author} {\bibfnamefont {T.}~\bibnamefont
  {Fang}}, \bibinfo {author} {\bibfnamefont {A.}~\bibnamefont {Konar}},
  \bibinfo {author} {\bibfnamefont {H.}~\bibnamefont {Xing}}, \ and\ \bibinfo
  {author} {\bibfnamefont {D.}~\bibnamefont {Jena}},\ }\href {\doibase
  10.1103/PhysRevB.78.205403} {\bibfield  {journal} {\bibinfo  {journal} {Phys.
  Rev. B}\ }\textbf {\bibinfo {volume} {78}},\ \bibinfo {pages} {205403}
  (\bibinfo {year} {2008})}\BibitemShut {NoStop}%
\bibitem [{\citenamefont {Betti}\ \emph {et~al.}(2011)\citenamefont {Betti},
  \citenamefont {Fiori},\ and\ \citenamefont {Iannaccone}}]{Betti2007APL}%
  \BibitemOpen
  \bibfield  {author} {\bibinfo {author} {\bibfnamefont {A.}~\bibnamefont
  {Betti}}, \bibinfo {author} {\bibfnamefont {G.}~\bibnamefont {Fiori}}, \ and\
  \bibinfo {author} {\bibfnamefont {G.}~\bibnamefont {Iannaccone}},\ }\href
  {\doibase http://dx.doi.org/10.1063/1.3587627} {\bibfield  {journal}
  {\bibinfo  {journal} {Applied Physics Letters}\ }\textbf {\bibinfo {volume}
  {98}},\ \bibinfo {eid} {212111} (\bibinfo {year} {2011})}\BibitemShut
  {NoStop}%
\bibitem [{\citenamefont {Akhavan}\ \emph {et~al.}(2012)\citenamefont
  {Akhavan}, \citenamefont {Jolley}, \citenamefont {Umana-Membreno},
  \citenamefont {Antoszewski},\ and\ \citenamefont
  {Faraone}}]{PhononLimited2012APL}%
  \BibitemOpen
  \bibfield  {author} {\bibinfo {author} {\bibfnamefont {N.~D.}\ \bibnamefont
  {Akhavan}}, \bibinfo {author} {\bibfnamefont {G.}~\bibnamefont {Jolley}},
  \bibinfo {author} {\bibfnamefont {G.~A.}\ \bibnamefont {Umana-Membreno}},
  \bibinfo {author} {\bibfnamefont {J.}~\bibnamefont {Antoszewski}}, \ and\
  \bibinfo {author} {\bibfnamefont {L.}~\bibnamefont {Faraone}},\ }\href
  {\doibase http://dx.doi.org/10.1063/1.4764318} {\bibfield  {journal}
  {\bibinfo  {journal} {Journal of Applied Physics}\ }\textbf {\bibinfo
  {volume} {112}},\ \bibinfo {eid} {094505} (\bibinfo {year}
  {2012})}\BibitemShut {NoStop}%
\bibitem [{\citenamefont {Grassi}\ \emph {et~al.}(2013)\citenamefont {Grassi},
  \citenamefont {Gnudi}, \citenamefont {Imperiale}, \citenamefont {Gnani},
  \citenamefont {Reggiani},\ and\ \citenamefont
  {Baccarani}}]{ModeSpace2013APL}%
  \BibitemOpen
  \bibfield  {author} {\bibinfo {author} {\bibfnamefont {R.}~\bibnamefont
  {Grassi}}, \bibinfo {author} {\bibfnamefont {A.}~\bibnamefont {Gnudi}},
  \bibinfo {author} {\bibfnamefont {I.}~\bibnamefont {Imperiale}}, \bibinfo
  {author} {\bibfnamefont {E.}~\bibnamefont {Gnani}}, \bibinfo {author}
  {\bibfnamefont {S.}~\bibnamefont {Reggiani}}, \ and\ \bibinfo {author}
  {\bibfnamefont {G.}~\bibnamefont {Baccarani}},\ }\href {\doibase
  http://dx.doi.org/10.1063/1.4800900} {\bibfield  {journal} {\bibinfo
  {journal} {Journal of Applied Physics}\ }\textbf {\bibinfo {volume} {113}},\
  \bibinfo {eid} {144506} (\bibinfo {year} {2013})}\BibitemShut {NoStop}%
\bibitem [{\citenamefont {Li}\ \emph {et~al.}(2008)\citenamefont {Li},
  \citenamefont {Qian}, \citenamefont {Wu}, \citenamefont {Gu},\ and\
  \citenamefont {Duan}}]{PhysRevLett.100.206802}%
  \BibitemOpen
  \bibfield  {author} {\bibinfo {author} {\bibfnamefont {Z.}~\bibnamefont
  {Li}}, \bibinfo {author} {\bibfnamefont {H.}~\bibnamefont {Qian}}, \bibinfo
  {author} {\bibfnamefont {J.}~\bibnamefont {Wu}}, \bibinfo {author}
  {\bibfnamefont {B.-L.}\ \bibnamefont {Gu}}, \ and\ \bibinfo {author}
  {\bibfnamefont {W.}~\bibnamefont {Duan}},\ }\href {\doibase
  10.1103/PhysRevLett.100.206802} {\bibfield  {journal} {\bibinfo  {journal}
  {Phys. Rev. Lett.}\ }\textbf {\bibinfo {volume} {100}},\ \bibinfo {pages}
  {206802} (\bibinfo {year} {2008})}\BibitemShut {NoStop}%
\bibitem [{\citenamefont {Nakabayashi}\ \emph {et~al.}(2009)\citenamefont
  {Nakabayashi}, \citenamefont {Yamamoto},\ and\ \citenamefont
  {Kurihara}}]{PhysRevLett.102.066803}%
  \BibitemOpen
  \bibfield  {author} {\bibinfo {author} {\bibfnamefont {J.}~\bibnamefont
  {Nakabayashi}}, \bibinfo {author} {\bibfnamefont {D.}~\bibnamefont
  {Yamamoto}}, \ and\ \bibinfo {author} {\bibfnamefont {S.}~\bibnamefont
  {Kurihara}},\ }\href {\doibase 10.1103/PhysRevLett.102.066803} {\bibfield
  {journal} {\bibinfo  {journal} {Phys. Rev. Lett.}\ }\textbf {\bibinfo
  {volume} {102}},\ \bibinfo {pages} {066803} (\bibinfo {year}
  {2009})}\BibitemShut {NoStop}%
\bibitem [{\citenamefont {Wakabayashi}\ \emph {et~al.}(2010)\citenamefont
  {Wakabayashi}, \citenamefont {Sasaki}, \citenamefont {Nakanishi},\ and\
  \citenamefont {Enoki}}]{1468-6996-11-5-054504}%
  \BibitemOpen
  \bibfield  {author} {\bibinfo {author} {\bibfnamefont {K.}~\bibnamefont
  {Wakabayashi}}, \bibinfo {author} {\bibfnamefont {K.}~\bibnamefont {Sasaki}},
  \bibinfo {author} {\bibfnamefont {T.}~\bibnamefont {Nakanishi}}, \ and\
  \bibinfo {author} {\bibfnamefont {T.}~\bibnamefont {Enoki}},\ }\href
  {http://stacks.iop.org/1468-6996/11/i=5/a=054504} {\bibfield  {journal}
  {\bibinfo  {journal} {Science and Technology of Advanced Materials}\ }\textbf
  {\bibinfo {volume} {11}},\ \bibinfo {pages} {054504} (\bibinfo {year}
  {2010})}\BibitemShut {NoStop}%
\bibitem [{\citenamefont {Yamada}\ \emph {et~al.}(2008)\citenamefont {Yamada},
  \citenamefont {Yamakita},\ and\ \citenamefont {Ohno}}]{PhysRevB.77.054302}%
  \BibitemOpen
  \bibfield  {author} {\bibinfo {author} {\bibfnamefont {M.}~\bibnamefont
  {Yamada}}, \bibinfo {author} {\bibfnamefont {Y.}~\bibnamefont {Yamakita}}, \
  and\ \bibinfo {author} {\bibfnamefont {K.}~\bibnamefont {Ohno}},\ }\href
  {\doibase 10.1103/PhysRevB.77.054302} {\bibfield  {journal} {\bibinfo
  {journal} {Phys. Rev. B}\ }\textbf {\bibinfo {volume} {77}},\ \bibinfo
  {pages} {054302} (\bibinfo {year} {2008})}\BibitemShut {NoStop}%
\bibitem [{\citenamefont {Gillen}\ \emph {et~al.}(2009)\citenamefont {Gillen},
  \citenamefont {Mohr}, \citenamefont {Thomsen},\ and\ \citenamefont
  {Maultzsch}}]{PhysRevB.80.155418}%
  \BibitemOpen
  \bibfield  {author} {\bibinfo {author} {\bibfnamefont {R.}~\bibnamefont
  {Gillen}}, \bibinfo {author} {\bibfnamefont {M.}~\bibnamefont {Mohr}},
  \bibinfo {author} {\bibfnamefont {C.}~\bibnamefont {Thomsen}}, \ and\
  \bibinfo {author} {\bibfnamefont {J.}~\bibnamefont {Maultzsch}},\ }\href
  {\doibase 10.1103/PhysRevB.80.155418} {\bibfield  {journal} {\bibinfo
  {journal} {Phys. Rev. B}\ }\textbf {\bibinfo {volume} {80}},\ \bibinfo
  {pages} {155418} (\bibinfo {year} {2009})}\BibitemShut {NoStop}%
\bibitem [{\citenamefont {Ziman}(1960)}]{ziman}%
  \BibitemOpen
  \bibfield  {author} {\bibinfo {author} {\bibfnamefont {J.~M.}\ \bibnamefont
  {Ziman}},\ }\href@noop {} {\emph {\bibinfo {title} {Electrons and Phonons:The
  Theory of Transport Phenomena in Solids}}}\ (\bibinfo  {publisher} {Oxford
  University Press},\ \bibinfo {address} {Oxford},\ \bibinfo {year}
  {1960})\BibitemShut {NoStop}%
\end{thebibliography}%

\end{document}